\documentclass[aps,prl, twocolumn,superscriptaddress,floatfix,footinbib,10pt]{revtex4-2}
\usepackage{amsmath,amssymb,bm}
\usepackage{graphicx}
\usepackage{epstopdf}
\usepackage{latexsym}
\usepackage[normalem]{ulem} 
\usepackage[caption=false]{subfig}
\usepackage[usenames,dvipsnames]{color}
\usepackage[sectionbib]{bibunits}
\usepackage{hyperref}
\usepackage{physics}
\usepackage{natbib}
\usepackage{array}
\usepackage{bbold}
\usepackage{bm}
\usepackage{tabularx}
\usepackage{tikz}
\usepackage{multirow}
\usepackage{comment}
\usepackage{leftindex}

\defaultbibliographystyle{apsrev4-2} 
\defaultbibliography{RAIM_bibliography} 

\newcommand{\affA}{Van der Waals-Zeeman Institute, Institute of Physics, University of Amsterdam, 1098 XH Amsterdam, the Netherlands}
\newcommand{\affB}{QuSoft, Science Park 123, 1098 XG Amsterdam, the Netherlands}
\newcommand{\affC}{Institute for Theoretical Physics, Institute of Physics, University of Amsterdam, Science Park 904, 1098 XH Amsterdam, the Netherlands}

\makeatletter
\def\maketitle{
\@author@finish
\title@column\titleblock@produce
\suppressfloats[t]}
\makeatother

\begin{document}

\title{Vibrationally coupled Rydberg atom-ion molecules}
\date{\today}

\author{Ilango Maran}\affiliation{\affA}
\author{Liam J. Bond}\affiliation{\affB}\affiliation{\affC}
\author{Jeremy T. Young}\affiliation{\affC}
\author{Arghavan Safavi-Naini}\affiliation{\affB}\affiliation{\affC}
\author{Rene Gerritsma}\affiliation{\affA}\affiliation{\affB}

\begin{abstract} 
We study the occurrence of Rydberg atom-ion molecules (RAIMs) in a hybrid atom-ion system with an ion crystal trapped in a Paul trap coupled to Rydberg atoms on its either ends. To assess the feasibility of such a system, we perform a detailed Floquet analysis of the effect of the Paul trap's rf potential on the RAIMs and provide a qualitative analysis of the survival probability based on scaling laws. We conclude that the RAIM survives for sufficiently weak and low frequency traps. We then use this hybrid system and propose a scheme to utilise the common motional modes of the ion crystal to suppress (blockade) or enhance (anti-blockade) the probability of forming two RAIMs at the ends of the chain, replacing the typical blockade radius by the length of the ion crystal.
\end{abstract}

\maketitle  

Recently, a new type of Rydberg molecule consisting of an ion and a Rydberg atom was proposed and observed in an ultracold cloud of rubidium atoms~\cite{Duspayev_2021PRR,Deiss_2021Atoms,Zuber_2022Nat,Zou_2023PRL}. The binding mechanism is based on the flipping dipole induced in the Rydberg atom by the ion's electric field, which results in an atom-ion separation-dependent mixing of the different Rydberg states~\cite{Hahn_2000PRA,Secker_2016PRA,Secker_2017PRL,Engel_2018PRL,Haze_2019arXiv,Ewald_2019PRL,mudli2024ionmediatedinteractioncontrolledphase,Müller_2008}. Molecular bound states are found in the potential wells located at the avoided crossings between the low orbital angular momentum Rydberg state ($nS$ and $nP$) and the hydrogenic manifolds. These molecular ions can have remarkably large bond lengths on the order of a few $\mu$m and binding energies as low as a few MHz~\cite{Duspayev_2021PRR,Deiss_2021Atoms,Zuber_2022Nat,Zou_2023PRL}. 
\begin{figure}[h!]
 \includegraphics[width=\linewidth]{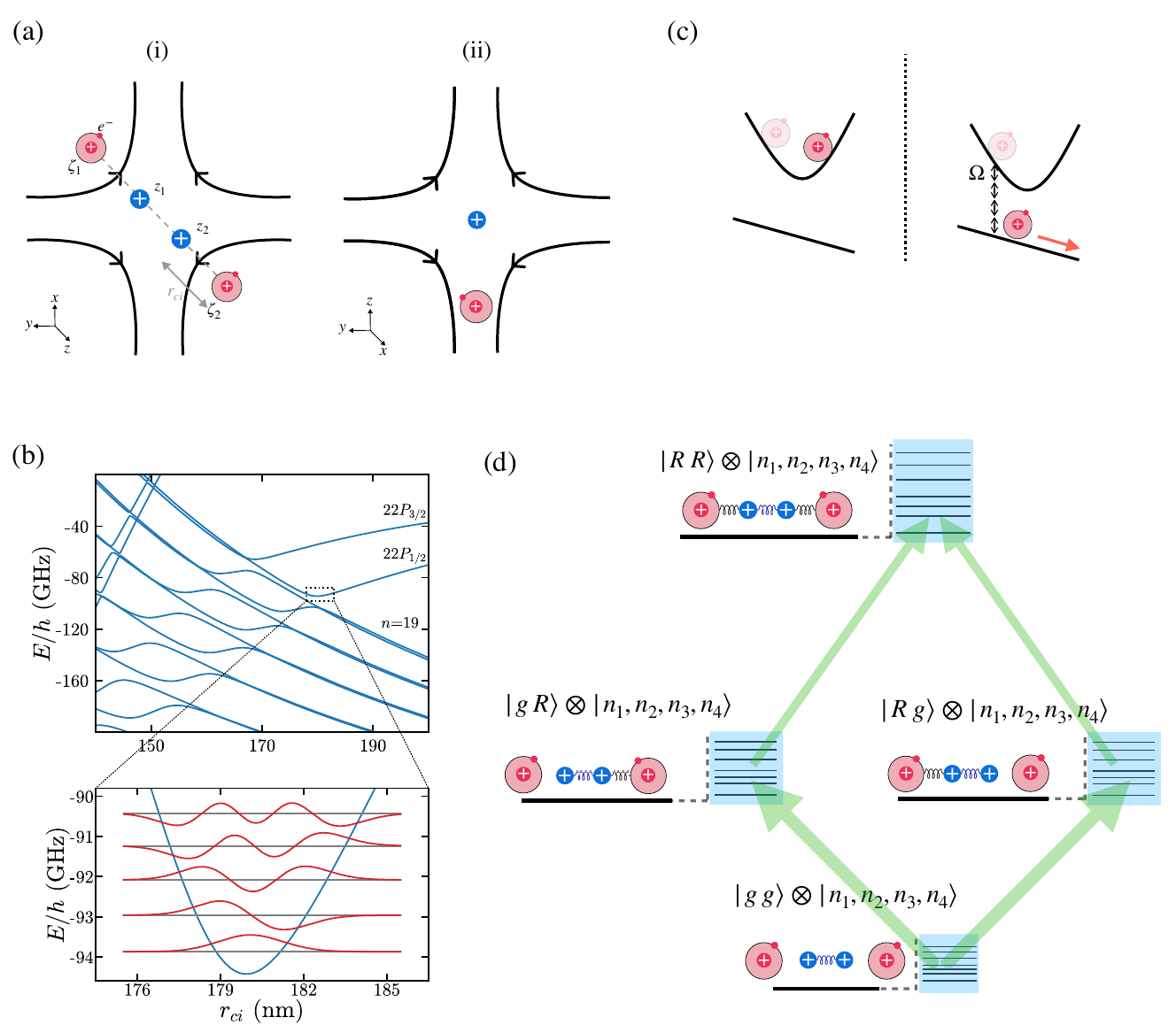}
 \caption{(a) Two vibrationally coupled RAIMs aligned along the rf-null axis of a Paul trap potential in the $xy$-plane (axial case) where $z_1$, $z_2$, $\zeta_1$ and $\zeta_2$ represent the positions of the ions and the cores of the Rydberg atoms respectively is shown in (i). A RAIM aligned along the $z$-axis of a Paul trap potential in the $zy$-plane (radial case) is shown in (ii). (b) Potential energy curves of a ${}^{87}{\rm Rb}{}^{9}{\rm Be}^+$ RAIM for $\vert m_j \vert= 1/2$ in the vicinity of the 22$P$ state is shown above and a magnified view of a part of the 22$P_{1/2}$ potential well with its first five bound states is shown below. The energies are relative to the field-free atomic 22$P_{3/2}$ state. (c) Schematic depicting a Rydberg atom escaping the well state by coupling with another state through a high-order photon process. (d) Energy levels of the four possible electronic states of the two atoms, each with its own ladder of motional states characterized by their respective phonon occupation numbers that can be exploited to mediate long-range Rydberg interactions.}
 \label{fig:fig1}
\end{figure}

A natural question is whether we can extend the Rydberg atom-ion molecule to include more particles. In this letter, we study the possibility of replacing the single ion with a trapped ion crystal. We show that the Rydberg atom-ion molecule (RAIM) survives the effects of the Paul trap, provided the trap is sufficiently weak and low-frequency, and utilize the rich vibrational spectrum added by the additional ions to propose novel longer-range Rydberg blockade and anti-blockade mechanisms. 

The structure of the paper is as follows. We illustrate the system in Fig.~\ref{fig:fig1}(a)(i), where a two ion chain interacts with Rydberg atoms placed at either end, to form two RAIMs. In Fig.~\ref{fig:fig1}(b), we plot the potential energy curves due to the Coulombic interaction between the ion and the Rydberg atom, assuming that the atomic core and the ion are stationary compared to the Rydberg electron~\cite{supp}. The potential wells support RAIMs at specific values of the core-ion separation $r_{\rm ci}$, one to three orders of magnitude smaller than the typical ion-ion separation in the crystal. Next, we use Floquet formalism to study the effect of the Paul trap's rf-drive on the RAIM, which can potentially induce transitions from the molecular state to unbound atomic states, as shown in Fig.~\ref{fig:fig1}(c). We use a simple qualitative model using scaling laws to understand the RAIM's survival probability and identify a broad range of Rydberg states and Paul trap drive frequencies where trap-induced losses will be negligible. Finally, we use the common motional modes of the trapped ion crystal to mediate interactions between the two Rydberg atoms. By resonantly driving specific transitions between phonon modes, we engineer an asymmetry in the level scheme, shown in Fig.~\ref{fig:fig1}(d), to realize either blockade or anti-blockade over a range set by the ion crystal's length rather than the shorter typical blockade radius set by the Rydberg dipole-dipole interaction. This showcases the potential of hybrid atom-ion systems as an excellent platform for quantum technology. 

\noindent\emph{Rydberg atom-ion molecule in a Paul trap:} 
We consider the case where a singly-charged ion of mass $m_{\rm i}$ is trapped in a Paul trap potential in the $zy$-plane interacting with a Rydberg atom (as in Fig.~\ref{fig:fig1}(a)(ii)). The position of the Rydberg atom's core defines the origin and the ion is at the Paul trap's rf-null, $z=r_{\rm ci}$. The Rydberg electron is at $r_{\rm e}$ and $r_{\rm ei}$ is the separation between the electron and the ion. Within the Born-Oppenheimer approximation, where the ion and the core are both considered stationary, the system is described by the time-dependent Hamiltonian~\cite{supp,ROBERTSON2021107814,van2009dipole}
\begin{equation}\label{eq:Htot}
\begin{split}
    H(r_{\rm ci},t)&=H_{\rm TI}(r_{\rm ci})+H_{\rm P,rad}(r_{\rm ci},t),\\
    H_{\rm TI}(r_{\rm ci}) &= H_0 + H_{\rm ai}(r_{\rm ci}).
\end{split}
\end{equation}
Here, $H_{\rm TI}$ is the time-independent Hamiltonian with $H_0=\frac{p_e^2}{2m_{\rm e}} + V(r_{\rm e})$, $V(r_{\rm e})$ is the binding potential of the Rydberg atom, and $H_{\rm ai}$ is the coulombic interaction between the Rydberg atom and the ion. We expand this interaction in the multipole series truncated at the 64$^{\rm th}$ pole ($l^\prime$= 6)   
\begin{equation}\label{eq: Hai}
  H_{\rm ai}(r_{\rm ci})= -\frac{e^2}{4\pi\epsilon_0}\sum_{l^\prime=1}^{6}\sqrt{\frac{4\pi}{2l^\prime+1}}\frac{r_{\rm e}^{l^\prime}}{r_{\rm ci}^{l^\prime+1}}Y_{l^\prime0}(\theta_{\rm e},\phi_{\rm e}),
\end{equation}
where $e$ is the elementary charge, $\epsilon_0$ is the vacuum permittivity and $Y_{l^\prime0}$ correspond to the spherical harmonics of degree $l^\prime$ and order $0$. We numerically diagonalize $H_{\rm TI}$ at a given value of $r_{\rm ci}$ to obtain the eigenvalues $\lbrace E_\alpha(r_{\rm ci})\rbrace$ and their respective eigenvectors $\lbrace \vert \psi_\alpha(r_{\rm ci})\rangle \rbrace$~\cite{supp}. Fig.~\ref{fig:fig1}(b) shows a typical spectrum analogous to the potential energy curves obtained in Ref.~\cite{Deiss_2021Atoms}. These curves can be interpreted as atom-ion molecular potentials, with the radial bound states of each of the wells in the spectrum corresponding to vibrational eigenstates of the atom-ion system at a particular angular momentum. In the following we will focus on states of one of these bound state wells and use $\vert \psi_{\rm RAIM}(r_{\rm ci})\rangle$ to indicate its eigenstate at a given value of $r_{\rm ci}$.
\begin{figure*}
    \centering    
    \includegraphics[width=7in]{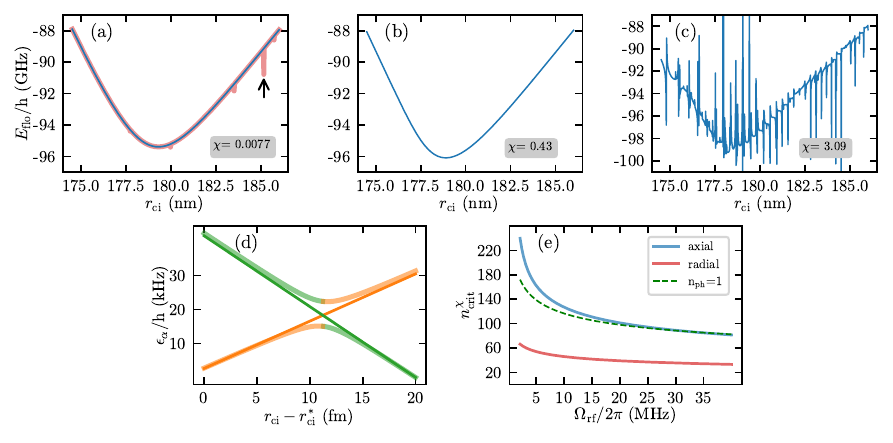}
    \caption{Energy expectation value of the static Hamiltonian $H_{\rm TI}$ for the Floquet state $\ket{\phi_{\rm RAIM}}$ ($E_{\rm flo}$) that correspond to the 22$P_{1/2}$ molecular well state of a ${}^{87}{\rm Rb}{}^{9}{\rm Be}^+$ RAIM oriented along one of the radial axis of a sinusoidal Paul trap which traps ${}^{9}{\rm Be}^+$ ions with $q = 0.1$ driven at (a) 2$\pi$ $\times$ 20 MHz (100 (red transparent) and 8000 (blue) Trotter steps), (b) 2$\pi$ $\times$ 150 MHz (8000 Trotter steps) and (c) 2$\pi$ $\times$ 400 MHz (8000 Trotter steps). Energies are relative to the field-free atomic 22${P}_{3/2}$ state. The scan over $r_{\rm ci}$ was done at a resolution of 10 pm. (d) Avoided crossing between the quasienergy of the Floquet state that corresponds to the well state (orange) and to the other eigenstate (green) it couples with, at the core-ion distance ($r_{\rm ci}$) marked by the black arrow ($r^{*}_{\rm ci}=185.15$ nm) in Fig. \ref{fig:Rb87_n=22_avoided crossings}(a) for a sinusoidal trap. The apparent avoided crossing for 100 Trotter steps (thick transparent lines) disappears for 8000 Trotter steps (thin solid lines). (e) $n^\chi_{\rm crit}$ plotted as a function of $\Omega_{\rm rf}$ for the RAIM oriented along the radial (red) and rf-null (blue) axis. The area below the green dashed curve indicates the region of parameter space that needs a multi-photon process for a LZ transition.}
    \label{fig:Rb87_n=22_avoided crossings}
\end{figure*}

Next, we characterize the trap drive's effect on the energy spectrum of $H_{\rm TI}$ and quantify the survival probability of the RAIM in the trap. The Hamiltonian $H_{\rm P,rad}(r_{\rm ci}, t)$ describes the Rydberg electron, Rydberg core, and the ion in the Paul trap potential which is given by~\cite{supp},
\begin{equation}
\label{eq:HPaul}
\begin{split} 
H_{\rm P,rad}(r_{\rm ci}, t)= &\frac{m_{\rm i}q\Omega_{\rm rf}^2}{4}\biggl(\frac{r_{\rm e}^2\sin^2{\theta_{\rm e}}}{2}-\frac{r_{\rm e}^2\sin^2{\theta_{\rm e}}\cos{2\phi_{\rm e}}}{2}\\&-r_{\rm e}^2\cos^2{\theta_{\rm e}}+2r_{\rm ci}r_{\rm e}\cos{\theta_{\rm e}}\biggr) \cos(\Omega_{\rm rf}t),
\end{split}
\end{equation}
where $q$ ($=|q_z|=|q_y|$) and $\Omega_{\rm rf}$ are the dynamic stability parameter (measure of the electric field gradient in the trap)\cite{RevModPhys.75.281} and the trap drive frequency respectively.

We use Floquet formalism to analyze the full Hamiltonian. The Floquet Hamiltonian at each $r_{\rm ci}$ is given by  $\mathcal H_{\mathcal F}=\frac{i\hbar}{T} \ln(\mathcal F_0)$ where $\mathcal F_0= \mathcal{T} \exp[\frac{-i}{\hbar}\int_{0}^{T} H(r_{\rm ci},\tau) \,d\tau]$ is the Floquet propagator ($\mathcal{T}$ denotes time-ordering) and $T=2\pi/\Omega_{\rm rf}$. The eigenvalues $\lbrace \epsilon_\alpha (r_{\rm ci})\rbrace$ and eigenvectors $\lbrace \vert \phi_\alpha (r_{\rm ci})\rangle \rbrace$ of $\mathcal H_{\mathcal F}$ are the quasienergies and Floquet modes respectively. The rf-drive frequency sets the width of the band of Floquet quasienergies.  

We restrict ourselves to a subspace of the Rydberg atom's internal states characterized by $n$, $s$, $l$, $j$, and fix $m_j$ to one of the two values $\pm 1/2$. We note that the Hamiltonian $H_{\rm 
P,rad}$, with the exception of the second term which for $n=22$ is a factor of $\sim 1000$ times smaller than the dominant term $2 r_{\rm ci} r_{\rm e} \cos \theta_{\rm e}$, does not couple sectors with different $m_j$ values.

In general, if the Paul trap is a perturbation on the static terms, we can find one Floquet state with significant overlap with each eigenvector in $\lbrace \vert \psi_\alpha(r_{\rm ci})\rangle \rbrace$, which also includes $\vert \psi_{\rm RAIM}(r_{\rm ci})\rangle$. We use this procedure to scan over a range of $r_{\rm ci}$ and at each point identify the Floquet state $\vert \phi_\alpha (r_{\rm ci})\rangle$ with significant overlap with $\vert \psi_{\rm RAIM}(r_{\rm ci})\rangle$, which we call $\vert \phi_{\rm RAIM} (r_{\rm ci})\rangle$. Due to the relatively small values of $\Omega_{\rm rf}$ compared to the Rydberg energy scales, the energy spectrum is highly compressed in the quasienergy spectrum, inhibiting a clear physical interpretation of their values. To circumvent this issue, we evaluate $E_{\rm flo}=\langle \phi_{\rm RAIM} \vert H_{\rm TI} \vert \phi_{\rm RAIM} \rangle$, which provides a useful picture for visualizing the role of couplings to other Rydberg states due to the Paul trap, see Fig.~\ref{fig:Rb87_n=22_avoided crossings}. 

In Fig.~\ref{fig:Rb87_n=22_avoided crossings}(a), (b) and (c), we numerically calculate $E_{\rm flo}$ using first order Trotter decomposition, in which we approximate the sinusoidal drive by small time steps of constant amplitude. The emergences of spikes in the energy herald the presence of avoided crossings in the quasienergies. Physically, these avoided crossings correspond to the absorption and emission of $\text{n}_{\rm ph}$ photons of frequency $\Omega_{\rm rf}$, which may cause the RAIM to dissociate (illustrated schematically in Fig.~\ref{fig:fig1}(c)). As we increase the number of Trotter steps, the vast majority of these avoided crossings disappear like the example in Fig.~\ref{fig:Rb87_n=22_avoided crossings}(d), which we associate with fictitous frequency components in the drive that emerge due to Trotterization.

While we expect the presence of the avoided crossings to reduce the RAIM's lifetime, we need to quantify how the increasing number of crossings with the trap drive (as seen in Fig.~\ref{fig:Rb87_n=22_avoided crossings}(a), (b) and (c)), which reduces the number of photons (n$_{\rm ph}$) to couple to a nearby state and thus enhances its strength, affects the survival probability of the RAIM. Note that a fully numerical analysis becomes impractical for larger principal quantum number $n$, as the coupling between Rydberg states with different $m_j$ values start becoming significant requiring a prohibitively large Hilbert space. 

Instead, we identify parameter regimes where the RAIM survives, using an approach based on Landau-Zener-St\"{u}ckleberg interferometry which studies the effect of an oscillating field (e.g., the Paul trap here) on Landau-Zener (LZ) processes \cite{SHEVCHENKO20101}, aided by scaling arguments. To identify the critical parameters at which the RAIM no longer survives in the Paul trap, we consider an effective two-level treatment. This is well-justified by the fact that for relevant parameter regimes, the corresponding excitations involve multi-photon resonances, in which case the onset of LZ transitions will be dominated by the nearest state (in energy) which requires the least number of photons. This is illustrated by the fact that most of the resonances in Fig.~\ref{fig:Rb87_n=22_avoided crossings}(c) are equally spaced, corresponding to different numbers of photons for the same level. 

The energy gap between the well state and the nearest (in energy) Rydberg state $K$ at the location of the well minima is $\Delta E \approx 0.28 E_{\rm H}/n^4$, where $E_{\rm H}$ is the Hartree energy \cite{supp}. The extra power of $n$ compared to the usual Rydberg energy separations ($\sim E_{\rm H}/n^3$) comes from the conservation of $m_J$ and the fact that each Stark fan contains $n$ different angular momentum ($l$) states \cite{gallagher1994rydberg}, while the prefactor is largely determined by the Rydberg atom's quantum defects. Since each photon has energy $\hbar \Omega_{\rm rf}$, the minimum number of photons needed for a LZ transition is $\text{n}_{\rm ph} \equiv \Delta E/\hbar \Omega_{\rm rf}$. Thus, we find a principal quantum number $n=0.727(E_{\rm H}/(\text{n}_{\rm ph}\hbar \Omega_{\rm rf}))^{1/4}$ which can be used to indicate regions of parameter space ($n$, $\Omega_{\rm rf}$) which need more than a single photon for a LZ transition, as in Fig.~\ref{fig:Rb87_n=22_avoided crossings}(e).

When $\text{n}_{\rm ph} \gtrsim 1$, single-photon processes become less important, and the probability of a LZ transition in each period is increasingly unlikely when $g \lesssim \Delta E$ \cite{SHEVCHENKO20101}, so we take $\chi=g/\Delta E \approx 1$ to define critical parameter values below which the RAIM is likely to survive. However, we note that parameters close to these critical values (particularly for few-photon processes) may warrant a more careful analysis to precisely determine the loss rates. While destructive interference for $\chi \gtrsim 1$ or resonances which occur at non-integer numbers of photons can reduce the probability of LZ transitions, the wavefunction of the RAIM spans many transition frequencies, so there will always be important LZ transitions when $\chi \gtrsim 1$.

When the RAIM is formed along one of the trap's radial axes (as in Fig.~\ref{fig:fig1}(a)(ii)), the coupling strength between the well state and the atomic state $K$ is dominated by the last term in Eq.~(\ref{eq:HPaul}) and is given by $g_{K, {\rm rad}}= m_{\rm i}q\Omega_{\rm rf}^2dz_{{\rm e},K}/2$. Here, $d$ represents the core-ion distance ($r_{\rm ci}$) at the minimum of the $nP_{1/2}$ molecular well state of the RAIM that scales like $d\approx 1.85 a_0 n^{5/2}$\cite{supp} as a consequence of the so-called Inglis-Teller limit \cite{gallagher1994rydberg}, where $a_0$ is the Bohr radius. We use the numerically obtained solutions at different $n$ to extract the scaling law for the transition dipole moment $ z_{{\rm e},K}\approx 0.6 a_0 n^{2}$. Thus, we find $g_{K,{\rm rad}}= 0.55m_{\rm i}q\Omega_{\rm rf}^2a_0^2n^{9/2}$, which gives a critical principal quantum number $n^{\chi}_{\rm crit,rad}= 0.923(E_{\rm H}/(m_{\rm i}q\Omega^2_{\rm rf}a_0^2))^{2/17}$. For Figs.~\ref{fig:Rb87_n=22_avoided crossings}(b, c), $\chi_{\rm rad}$ = 0.43, 3.09, respectively, so they fall on either side of the critical $\Omega_{\rm rf}$. Since the former has negligible crossings and the latter has significant crossings, the Floquet analysis is consistent with our scaling argument. 

When we consider the case where the RAIM is formed along the Paul trap's rf-null axis (as in Fig.~\ref{fig:fig1}(a)(i)), the coupling strength is given by $g_{K, {\rm ax}}= m_{\rm i}q\Omega_{\rm rf}^2\rho_{{\rm e},K}/4$, where $\rho_{{\rm e},K}=y_{{\rm e},K}^2-x_{{\rm e},K}^2$ scales as $4.8 a_0^2 n^{2.6}$. Thus, $g_{K,\rm ax}= 1.2 m_{\rm i}q\Omega_{\rm rf}^2a_0^2n^{2.6}$ which gives $n^{\chi}_{\rm crit,ax}= 0.8(E_{\rm H}/(m_{\rm i}q\Omega^2_{\rm rf}a_0^2))^{0.15}$. In this case, the coupling $g$ to the closest state is relatively small, so we consider the coupling due to the next-nearest states, which require several times as many photons for resonant excitations \cite{supp}, although we continue to use the smallest $\Delta E$ as a more stringent condition on the survivability.

\emph{Phonon-mediated (anti-) blockade:} Next, we show how we can use the dependence of the ion crystal's collective motional modes on whether there are zero, one, or two RAIMs at its ends to extend the range of Rydberg (anti-) blockade to the length of the ion crystal. We demonstrate our results using a crystal of two singly-charged ions of mass $m_{\rm i}$ at $z_1$ and $z_2$ with two atoms of mass $m_{\rm a}$ placed at the crystal's opposite ends at $\zeta_1$ and $\zeta_2$, respectively, such that $|\zeta_l - z_l|\approx d$ with $l=1,2$ indexing the ion-Rydberg pair forming the RAIM (see Fig.~\ref{fig:fig1}(a)(i)). Although we focus on the two ion case, our results can be generalised to longer ion crystals. We assume that the atoms are held in place using optical tweezers with a trap frequency $\omega_{\rm t}$. We approximate the RAIM potential near its minimum as harmonic with frequency $\omega_{\rm M}$. The ions are in a harmonic trap with tight transverse confinement, such that we may neglect the motion in this direction, and an axial trap frequency $\omega_{\rm i}$. Upon the application of a Rydberg excitation pulse using laser beams propagating along the trap's axis, the system can be in one of the four possible electronic states with zero, one, or two RAIMs. The potential energy of the crystal in the different electronic states is given by
\begin{equation}\label{eq: Potential energy}
    V_{\vert o \rangle } =\frac{1}{2}m_{\rm i}\omega_{\rm i}^2\left(z_1^2+z_2^2\right)-\frac{e^2}{4\pi\epsilon_0|z_1-z_2|}+U_{\vert o \rangle },
\end{equation}
where $o\in\lbrace o_1,o_2,o_3,o_4\rbrace=\lbrace gg,gR,Rg,RR\rbrace$,  $U_{\vert gg \rangle}= 0$, $U_{\vert R\, g\rangle}=U_{\vert g\, R\rangle}= \frac{1}{2}\mu\omega_{\rm M}^2(|z_l-\zeta_l|-d)^2$, and $U_{\vert R\, R \rangle}= U_{\vert g\, R\rangle}+ U_{\vert R\, g\rangle}$. We expect the RAIM's lateral motion to be too slow in comparison to the time scales in which we are interested ($\sim 10 \mu$s) to have any significant impact, and we neglect it in this work. We find the normal modes $\bm{v^{\vert o \rangle }}$ and corresponding mode frequencies $\bm{\omega^{\vert o \rangle}}$ for each of the four electronic states, as shown in Fig.~\ref{fig:fig1}(d)~\cite{Morigi:2001,Home:2013,supp}. For a given electronic state, the motional state is characterized by $N=(n_1,n_2,n_3,n_4)$ which represents the occupation numbers of the four modes in descending order of their mode frequencies. 

As an example, we consider a system with a crystal of two ${}^{9}{\rm Be}^+$ ions with an axial trap frequency $\omega_{\rm i}= 2\pi\times 1$ MHz and two ${}^{87}{\rm Rb}$ atoms placed at either ends at a distance $\sim d_{50P_{1/2}}$ ($\approx 1.735\mu$m) from the ion crystal using optical tweezers with trap frequency $\omega_{\rm t}= 2\pi\times 0.2$ MHz. We use a Rydberg pulse to excite to the $50P_{1/2}$ molecular well state of the ${}^{87}{\rm Rb}{}^{9}{\rm Be}^+$ RAIM, which has $\omega_{\rm M}= 2\pi\times 36$ MHz.

We prepare the system in the motional ground state of the trap, $\ket{\psi(0)} =\ket{gg,0000}$. For blockade, we resonantly drive transitions from $\ket{gg,0000}$ to $\ket{gR,0010}$ using a laser at a frequency $\omega_1$, as shown schematically in Fig.~\ref{fig:fig1}(d). Although there are a multitude of levels in the $\ket{RR}$ manifold that could potentially resonantly couple to $\ket{gR,0010}$, their overlaps are small, and thus $\ket{RR}$ remains unpopulated. In this way, we exploit both the existence and long-range nature of the motional modes $\ket{gR,N}$ to realize an effective asymmetry in the level spectrum. In Fig.~\ref{fig:blockadefacilitation}(a) we demonstrate this blockade mechanism by simulating the dynamics under
\begin{equation}\label{eq:Hblockadefacilitation}
\begin{split}
H&=H_{\rm d}+H_{\rm nd}, \\
H_{\rm nd}&=\frac{\hbar}{2}\sum_{N,N'}\Omega_{L}\big(\hat{\Sigma}_{1,NN'}+ \hat{\Sigma}_{2,NN'}\big)+h.c.,\\
H_{\rm d} &= \hbar\sum_{o,N}\omega_{o,N}\ket{o,N}\bra{o,N},
\end{split}
\end{equation}
where $\Omega_L=\Omega_1 e^{i\omega_1t}$ and the operators describing the transitions to Rydberg states are given by
\begin{align}
\hat{\Sigma}_{1,NN'}&=S^{o_1o_2}_{NN'}|gg,N\rangle\langle gR,N'|+S^{o_2o_3}_{NN'}|gg,N\rangle\langle Rg,N'|,\nonumber\\
\hat{\Sigma}_{2,NN'}&=S^{o_2o_4}_{NN'}|gR,N\rangle\langle RR,N'|+S^{o_3o_4}_{NN'}|Rg,N\rangle\langle RR,N'|.\nonumber
\end{align}
Here, the prefactors represent the wavefunction overlap integral between the Fock state $N$ of the electronic state $o_a$ and the Fock state $N'$ of the electronic state $o_b$ when the system is subjected to a Rydberg excitation pulse using a laser beam propagating along the system's axis, i.e., $S^{o_ao_b}_{NN'}=\langle o_a,N|e^{ik\hat{\zeta}_l'}|o_b,N'\rangle$ where $k$ is the (effective) wavenumber of the Rydberg laser. 
\begin{figure}
  \centering
  \includegraphics[width=\linewidth]{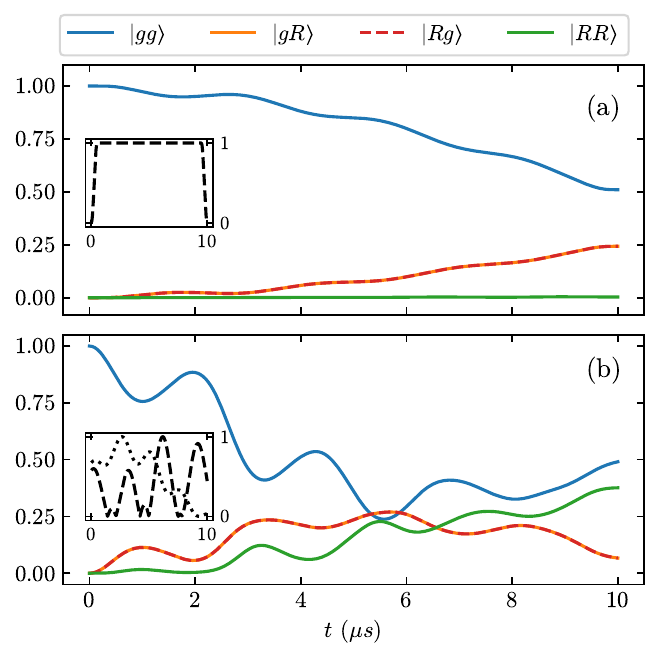}
  \caption{Population dynamics of $\ket{gg}$ (blue), $\ket{gR}$ (orange), $\ket{Rg}$ (dashed red) and $\ket{RR}$ (green). (a) We demonstrate blockade by using only a single driving field at $\omega_1$ with a square pulse shape (inset). (b) Dynamics of the anti-blockade protocol which uses two drive fields with their optimised pulse shapes (inset) at frequencies $\omega_1$ (dotted) and $\omega_2$ (dashed).}
 \label{fig:blockadefacilitation}
\end{figure} 
We set a Rabi frequency $\Omega_1 = 2\pi \times 0.057$ MHz and simulate evolution under a square pulse. The Rabi frequency was chosen based on the wavefunction overlap integrals to complete the state preparation in 10 $\mu $s which is well within the lifetime of the RAIM $\tau \approx 100 \mu$s (the radiative lifetime of 50$P$ state at 300 K)~\cite{Duspayev_2021PRR}. At $t_{\rm f} = 10 \mu $s, our protocol prepares the entangled state $\psi(t_{\rm f}) \approx 1/\sqrt{2} \ket{gg} + 1/2( \ket{gR} + \ket{Rg})$ with the maximum population of $\ket{RR}$ during the protocol ${\rm max}_{0\leq t \leq t_{\rm f}}[\abs{\braket{RR}{\psi(t)}}^2] = 0.004$. We therefore realize Rydberg blockade with a longer-range set by the ion crystal's length. 

To realize anti-blockade, we use an additional laser at frequency $\omega_2$ with strength $\Omega_2$, i.e., $\Omega_L = \Omega_1 f_1(t) e^{i \omega_1 t} + \Omega_2 f_2(t) e^{i \omega_2 t}$ ($f_{1,2}(t)$ are the pulse shapes), to resonantly drive the transition between $\ket{gR,0010}$ (and $\ket{Rg,0010}$) and $\ket{RR,0001}$. We choose to drive to $\ket{RR,0001}$ due to its high overlap with $\ket{gR,0010}$ and $\ket{Rg,0010}$. We use Rabi frequencies $\Omega_1 = 2\pi \times 0.153$~MHz and $\Omega_2 = 2\pi \times 0.14$~MHz. To minimize the impact of off-resonant coupling to levels in the $\ket{gR}, \ket{Rg}$ manifolds whilst still using a sufficiently large Rabi frequency to reduce the protocol time $t_{\rm f}$, we numerically optimize the pulse shapes  $f_i(t)$ of the driving fields~\cite{PhysRevLett.106.190501,supp}. In Fig.~\ref{fig:blockadefacilitation}(b), we plot the dynamics (colors) evolving under Eq.~(\ref{eq:Hblockadefacilitation}) and the optimised pulse shapes (black lines). 
At $t_{\rm f} \approx 10\mu$s, we prepare an entangled state of $\ket{gg}$ and $\ket{RR}$ such that the populations $|\braket{gR}{\psi(t_{\rm f})}|^2 = |\braket{Rg}{\psi(t_{\rm f})}|^2 = 0.07$ are small.

In summary, we showed that the recently observed molecular bound state between a Rydberg atom and an ion can survive in the presence of the rf-drive of a Paul trap. We presented an extensive numerical study, utilizing the Floquet formalism, to demonstrate that while strong trap drives may cause the RAIM to dissociate, the molecule remains stable for an extended region of parameter space. Furthermore, we presented an analytic scaling relation which can be applied as a rule of thumb to a variety of systems to determine their stability regimes. Finally, we devised a protocol utilizing the modification to the collective motion of the trapped ion crystal due to the formation of the RAIM on the ends of an arbitrary length crystal to generate phonon-mediated, long-ranged, Rydberg (anti-)blockade. Our analysis presents a new exploratory direction for hybrid Rydberg-ion quantum simulation platforms, where the scalability of currently available neutral atom platforms and the long range connectivity of ion chains can be combined. The ion chain effectively acts as a bus that transports quantum information between two (ensembles of) neutral atoms, and it could be viewed as a viable alternative to the tweezer shuttling techniques used for long range connectivity in neutral atom platforms. Moreover, the quantized motion in the collective motional modes can be used to simulate quantum many body systems involving bosonic degrees of freedom, such as spin-boson models. 

\begin{acknowledgments}
This work was supported by the Netherlands Organization for Scientific Research (Grant Nos. 680.91.120 and VI.C.202.051, (R.G.). A.S.N is supported by the Dutch Research Council (NWO/OCW), as part of the Quantum Software Consortium programme (project number 024.003.037) and Quantum Delta NL (project number NGF.1582.22.030). J.T.Y.~was supported by the NWO Talent Programme (project number VI.Veni.222.312), which is (partly) financed by the Dutch Research Council (NWO).
\end{acknowledgments}

\bibliography{RAIM_bibliography}


\clearpage

\setcounter{secnumdepth}{1}
\setcounter{equation}{0}
\setcounter{figure}{0}
\setcounter{table}{0}
\setcounter{page}{1}

\renewcommand\thefigure{\arabic{figure}}

\let\theequationWithoutS\theequation 
\renewcommand\theequation{S\theequationWithoutS}
\let\thefigureWithoutS\thefigure 
\renewcommand\thefigure{S\thefigureWithoutS}

\renewcommand*{\citenumfont}[1]{S#1}
\renewcommand*{\bibnumfmt}[1]{[S#1]}

\title{Supplemental Material: Vibrationally coupled Rydberg atom-ion molecules}

\maketitle

\begin{bibunit}

\onecolumngrid

The supplemental material is organized as follows: In Sec.~\ref{sec:potential}, we discuss how to calculate the potential energy curves and eigenstates of the RAIMS. In Sec.~\ref{sec:Phonon modes}, we show how the ion chain mediates interactions via the motional modes. In Sec.~\ref{sec:overlap}, we calculate the overlap integrals for both the atomic states and the motional Fock states. In Sec.~\ref{sec:Paul}, we derive the Hamiltonian for a Paul trap in the two configurations of the atom and molecule discussed in the main text. In Sec.~\ref{sec:Effect of the second ion and axial trap on RAIM}, we discuss the effects of the second ion and the axial trap electrodes for two possible orientations of the RAIM along the rf-null axis. In Sec.~\ref{sec:scaling}, we present the details of the numerical scaling laws used in the main text to determine the survivability of the RAIMs. In Sec.~\ref{sec:digital}, we discuss the survivability of the RAIMs in a digital Paul trap using an argument based on the Landau-Zener criterion. Finally, in Sec.~\ref{sec:pulse}, we discuss the optimization of the pulse profiles used for entangling the atoms via anti-blockade.

\section{Potential energy curves of Rydberg atom-ion molecules}
\label{sec:potential}
In this section, we explain the procedure used to calculate the molecular potential energy curves of RAIMs~\cite{Duspayev_2021PRR,Deiss_2021Atoms}. As discussed in the main text, the molecular potentials of the RAIMs (i.e.,~Stark maps of the Rydberg atom in the presence of a nearby ion) are a result of the Coulombic interaction between the ion and the Rydberg atom. Here, we consider a neutral ${}^{87}{\rm Rb}$ atom interacting with a ${}^{9}{\rm Be}^+$ ion. We assume that the Rydberg atom's ionic core and the ion are both stationary and that the ionic core is located at the origin with the ion on the $z$-axis at $r_{\rm ci}$. The Stark maps of the Rydberg atom can now be calculated using the Hamiltonian
\begin{align}
  \label{eqS:Htot}
     H\left(r_{\rm ci}\right) &= H_0 + H_{\rm ai}\left(r_{\rm ci}\right),
\end{align}
where $H_0$ represents the Hamiltonian of the field-free Rydberg atom including the fine structure and ignoring the hyperfine structure and $H_{\rm ai}$ represents the Coulomb interaction between the ion and the Rydberg atom. The eigenstates of $H_0$ are the atomic energy levels characterized by $n$, $s$, $l$, $j$ and $m_j$. The corresponding eigenenergies of these eigenstates ($\ket{n,s,l,j,m_j}$) are obtained using the ARC (Alkali.ne Rydberg Calculator) package~\cite{ROBERTSON2021107814}. Analytical expressions for the radial wavefunction of these eigenstates are given by modifications to the radial hydrogenic wavefunctions using quantum defect theory as~\cite{van2009dipole}
\begin{align}
  \label{eqS:Radial wavefunction}
     R_{n^*l^*}(r_{\rm e})&=\sqrt{\bigg(\frac{2}{n^*}\bigg)^3\frac{(n^*-l^*-1)!}{(n^*+l^*)!(2n^*)}}e^{-r_{\rm e}/n^*}\bigg(\frac{2r_{\rm e}}{n^*}\bigg)^{l^*} L_{n^*-l^*-1}^{2l^*+1}\bigg(\frac{2r_{\rm e}}{n^*}\bigg),
\end{align}
where $n^*= n - \delta_{nlj}$, $l^*= l - \delta_{nlj} + I(l)$, and $\delta_{nlj}$ is the quantum defect which is mainly determined by $l$. The integer $I(l)$ fixes the number of radial nodes and is constrained to $\delta_{nlj} - l - 1/2<I(l) \le n_{\rm min} - l - 1$, where $n_{\rm min}$ is the principal quantum number of the ground state. For ${}^{87}{\rm Rb}$, we use $I(l)=\lfloor \delta_{nlj} \rfloor$. $L^{2l^*+1}_{n^*-l^*-1}$ are associated Laguerre polynomials.

We treat the Rydberg atom as an ionic core and a Rydberg electron, so $H_{\rm ai}$ is given by the Coulomb interaction energy between three point charges 
\begin{equation}\label{eqS:Hcou}
\begin{split}
     H_{\rm ai} (r_{\rm ci}) &= \frac{e^2}{4\pi\epsilon_0}\bigg(\frac{1}{|r_{\rm ci}|}-\frac{1}{|r_{\rm e}-r_{\rm ci}|}\bigg)\\
     &= -\frac{e^2}{4\pi\epsilon_0}\sum_{l^\prime=1}^{\infty}\sqrt{\frac{4\pi}{2l^\prime+1}}\frac{r_{\rm e}^{l^\prime}}{r_{\rm ci}^{l^\prime+1}}Y_{l^\prime0}(\theta_{{\rm e}},\phi_{{\rm e}}),
\end{split}
\end{equation}
where $e$ is the elementary charge, $\epsilon_0$ is the vacuum permittivity, $Y_{l^\prime0}$ is the spherical harmonic of degree $l^\prime$ and order $0$, and  ($r_{\rm e}$, $\theta_{\rm e}$, $\phi_{\rm e}$) is the Rydberg electron's position in spherical coordinates. 

\begin{figure}
    \centering
    \includegraphics{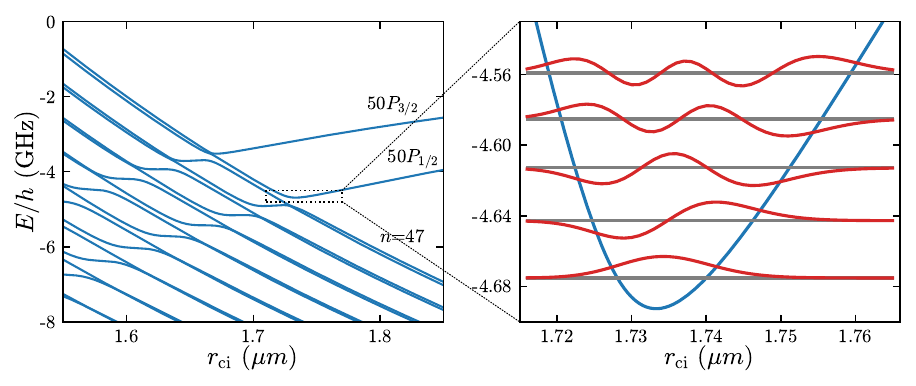}
    \caption{Potential energy curves of a ${}^{87}{\rm Rb}{}^{9}{\rm Be}^+$ RAIM for $\vert m_j \vert= 1/2$ in the vicinity of the 50$P$ state is on the left and a magnified view of a part of the 50$P_{1/2}$ potential well with its first five bound states is shown below. The energies are relative to the field-free atomic 50$P_{3/2}$ state.}
    \label{fig:Rb87_50P well}
\end{figure}

To diagonalize $H$ we first truncate the basis states by introducing an upper and a lower limit in the principal quantum number $n-4<n<n+4$. Next, we pick a specific $r_{\rm ci}$ and diagonalize $H\left(r_{\rm ci}\right)$ in this subspace to find the eigenstates and their corresponding eigenenergies. We repeat the procedure over a range of $r_{\rm ci}$ and plot the energy levels of the perturbed ${}^{87}{\rm Rb}$ atom in the vicinity of different $nP$ states (see Fig.~\ref{fig:Rb87_50P well}). The eigenstates are an admixture of the bare Rydberg states. 

Finally, we note that the matrix elements of $H\left(r_{\rm ci}\right)$ in the truncated subspace can be obtained analytically as
\begin{equation} \label{eqS:Hmatr}
  \begin{split}
     H\left(r_{\rm ci}\right) &= \bra{\psi}H_0|\Tilde{\psi}\rangle+ \bra{\psi}H_{\rm ai}\left(r_{\rm ci}\right)|\Tilde{\psi}\rangle,
     \end{split}
\end{equation}
where $\ket{\psi}= \ket{n,s,l,j,m_j}$ and $|\Tilde{\psi}\rangle= \vert\Tilde{n},\Tilde{s},\Tilde{l},\Tilde{j},\Tilde{m_j}\rangle$ represent the bare Rydberg basis states. We truncate the multipole expansion at the 64$^{th}$ pole ($l^\prime = 6$) and find 
\begin{equation}\label{eqn:second matrix 6 terms all mj}
\begin{split}
\langle\psi\vert H_{\rm ai}\left(r_{\rm ci}\right)\vert\Tilde{\psi}\rangle=&-\frac{e^2}{4\pi\epsilon_0}\langle n,s,l,j,m_{j}\vert\sum_{l^\prime=1}^{6}\frac{r_{\rm e}^{l^\prime}}{r_{\rm ci}^{l^\prime+1}}\sqrt{\frac{4\pi}{2l^\prime+1}}Y_{l^\prime 0}\vert \Tilde{n},\Tilde{s},\Tilde{l},\Tilde{j},\Tilde{m_{j}}\rangle\\
=&-\frac{e^2}{4\pi\epsilon_0}\sum_{l^\prime=1}^{6}\frac{1}{r_{\rm ci}^{l^\prime+1}}\bigg(\int_{0}^{\infty} R_{n^*l^*}(r_{\rm e}) r^{l^\prime+2}_{\rm e}R_{\Tilde{n}^*\Tilde{l}^*}(r_{\rm e})\,dr_{\rm e} \bigg)\bigg(\bra{l,s,j,m_{j}}\sqrt{\frac{4\pi}{2l^\prime+1}}Y_{l^\prime0}|\Tilde{l},\Tilde{s},\Tilde{j},\Tilde{m_{j}}\rangle\bigg)\\
=&-\frac{e^2}{4\pi\epsilon_0}\sum_{l^\prime=1}^{6}\frac{R_{\rm mat}A_{\rm mat}}{r_{\rm ci}^{l^\prime+1}}, 
\end{split}
\end{equation}
where $R_{\rm mat}$ are the radial matrix elements which are calculated numerically using the wavefunctions in Eq.~(\ref{eqS:Radial wavefunction}) and $A_{\rm mat}$ are the angular matrix elements given by, 
\begin{equation}\label{eqS: Ang int mj}
 A_{\rm mat}=(-1)^{2l+\frac{1}{2}+l^\prime+j+\Tilde{j}-m_{j}}\sqrt{(2l+1)(2\Tilde{l}+1)(2j+1)(2\Tilde{j}+1)}\begin{pmatrix}
l & l^\prime & \Tilde{l}\\
0 & 0 & 0
\end{pmatrix}\begin{pmatrix}
j & l^\prime & \Tilde{j}\\
-m_{j} & 0 & \Tilde{m}_{j}
\end{pmatrix}\begin{Bmatrix}
l & j & 1/2\\
\Tilde{j} & \Tilde{l} & l^\prime
\end{Bmatrix}. 
\end{equation}

\section{Collective motional modes}\label{sec:Phonon modes}
We consider a crystal of two singly-charged ions of mass $m_{\rm i}$ at $z_1$ and $z_2$ trapped in a Paul trap which is approximated to be harmonic with an axial trap frequency $\omega_{\rm i}$. We neglect the motion in the tightly confined transverse direction. We include two atoms of mass $m_{\rm a}$ positioned at each end of the crystal at $\zeta_1$ and $\zeta_2$, roughly a distance $d$ from the ions. Here, $d$ represents the core-ion distance ($r_{\rm ci}$) at the minimum of the anharmonic molecular potential curve of the RAIM that we intend to excite. We approximate the RAIM's potential near its minimum as harmonic with frequency $\omega_{\rm M}$, and assume that the atoms are held in place using optical tweezers with a trap frequency $\omega_{\rm t}$. 
When the four particle system is subjected to the necessary Rydberg excitation pulse using a laser beam propagating along its axis, the system can be in one of four possible electronic states: either zero, one (at each possible end of the chain), or two RAIMs. The potential energy of the crystal depends on whether a RAIM has formed, and is given by
\begin{equation}\label{eqS: Potential energy}
    V_{\vert o \rangle } =\frac{1}{2}m_{\rm i}\omega_{\rm i}^2\left(z_1^2+z_2^2\right)-\frac{e^2}{4\pi\epsilon_0|z_1-z_2|}+U_{\vert o \rangle },
\end{equation}
where $o\in\lbrace o_1,o_2,o_3,o_4\rbrace=\lbrace gg,gR,Rg,RR\rbrace$, $U_{\vert gg \rangle}= 0$, and $U_{\vert R\, g\rangle}=U_{\vert g\, R\rangle}= \frac{1}{2}\mu\omega_{\rm M}^2(|z_l-\zeta_l|-d)^2$, with $l= 1,2$ indexing the ion-Rydberg pair forming the RAIM, and  $U_{\vert R\, R \rangle}= U_{\vert g\, R\rangle}+ U_{\vert R\, g\rangle}$. The set of equilibrium positions of the four particles $\{z_{\vert o \rangle}^{\rm eq}\}$ in the different configurations can be found by solving the set of equations $\partial V_{\vert o \rangle}/\partial r_j= 0$, where $r=(z_1,z_2,\zeta_1,\zeta_2)$. With the second order being the leading term in the Taylor expansion of the potential about the equilibrium position, the mass-weighted Hessian matrix can be represented as~\cite{Morigi:2001,Home:2013}
\begin{equation}\label{eqS: Hessian}
    A_{\vert o \rangle}^{(jk)} =\frac{1}{\sqrt{m_j m_k}}\frac{\partial^2V_{\vert o \rangle}}{\partial r_j \partial r_k}\bigg\vert_{\big\{z_{\vert o \rangle}^{\rm eq}\big\}}.
\end{equation}
At this point it is convenient to switch to mass-weighted coordinates $r'=\big(\sqrt{m_{\rm i}}z_1,\sqrt{m_{\rm i}}z_2,\sqrt{m_{\rm a}}\zeta_1,\sqrt{m_{\rm a}}\zeta_2\big)$. The small displacements of the particles about their equilibrium position in this new coordinate system is given by $r''=(z''_1, z''_2, \zeta''_1, \zeta''_2)$. The normal modes $v^{{\vert o \rangle}}$ and their corresponding mode frequencies $\omega^{{\vert o \rangle}}$ are the eigenvectors and the square root of the eigenvalues of the Hessian respectively. 

If no RAIMs are created ($\vert g\, g \rangle$), we recover the normal modes of an ion crystal of two ions and the tweezer modes of the two ground state atoms with mode frequencies $\omega^{\vert g\, g \rangle}=\big\lbrace \sqrt{3}\omega_{\rm i}, \omega_{\rm i}, \omega_{\rm t}, \omega_{\rm t}\rbrace$. When one RAIM is created ($\vert R\, g \rangle$ or $\vert g\, R \rangle$), there is a lengthy analytic expression for the mode frequencies. In the limit $\omega_{\rm M}\gg\omega_{\rm i}$, the three collective mode frequencies are $\omega^{\vert R\, g \rangle}_1=\omega_{\rm M}+ (1-\beta)\omega_{\rm i}^2/\omega_{\rm M}\approx \omega_{\rm M}$ and $\omega^{\vert R\, g \rangle}_{2,3} =\omega_{\rm i}\sqrt{1+\beta\pm\sqrt{1+\beta(\beta-1)}}$, where $\beta=\frac{m_{\rm i}}{m_{\rm i}+m_{\rm a}}$ and $\omega^{\vert R\, g \rangle}=\omega^{\vert g\, R \rangle}$. The last two frequencies correspond to the eigenmodes of a mixed ion crystal of mass $m_{\rm i}$ and $m_{\rm a}+m_{\rm i}$. There is one ground state atom in either case which does not participate in the collective motion and vibrates at the tweezer frequency $\omega_{\rm t}$ ($=\omega^{\vert R\, g \rangle}_{4}$). When two RAIMs are created ($\vert R\, R \rangle$), the frequencies of the four collective modes are $\omega^{\vert R\, R \rangle}= \frac{1}{\sqrt2}\sqrt{c\,\omega_{\rm i}^2+\omega_{\rm M}^2 \pm\sqrt{c^2\omega_{\rm i}^4+2c(1-2\beta)\omega_{\rm i}^2\omega_{\rm M}^2+\omega_{\rm M}^4}}$, where $c= 3$ or 1. In the limit $\omega_{\rm M}\gg\omega_{\rm i}$, we obtain the mode frequencies $\omega^{\vert R\, R \rangle}=\lbrace\omega_{\rm M}, \omega_{\rm M}, \sqrt{3 \beta}\omega_{\rm i},\sqrt{\beta}\omega_{\rm i}\rbrace$. The last two frequencies correspond to the eigenmodes of an ion crystal with mass $m_{\rm a}+m_{\rm i}$.

\section{Calculating wavefunction Overlap integrals }
\label{sec:overlap}
In this section, we explain in detail the calculation of the wavefunction overlap integral between a Fock state $N$ of the electronic state $o_a$ and a Fock state $N'$ of the electronic state $o_b$ when the system is subjected to a Rydberg excitation pulse using a laser beam propagating along the system's axis, i.e.,~$S^{o_a,o_b}_{{N,N'}}=\langle o_a,N|e^{ik\hat{\zeta_l'}}|o_b,N'\rangle$. Here, $l= 1,2$ indexes the ion-Rydberg pair forming the RAIM, $N= (n_1,n_2,n_3,n_4)$ represents the occupation numbers of the four modes of the system in descending order of their mode frequencies and $k$ is the (effective) wavenumber of the Rydberg laser. Note that the wavefunction overlap integral between two different Fock states of the same electronic state (i.e.,~$S^{oo}_{{N,N'}}$), and between a Fock state of the configuration with no RAIMs and a Fock state of the electronic state with two RAIMs (i.e.,~$S^{o_1o_4}_{{N,N'}}$) are both zero. 

The wavefunction of a given mass-weighted coordinate $x_p$ vibrating at a frequency $\omega_p$ in the oscillator's natural units where energy and the coordinate are measured in units of $\hbar\omega_p$ and $\sqrt{\frac{\hbar}{\omega_p}}$, respectively, is given by
\begin{equation}\label{eqS: QHO wavefunction}
    \psi_{a}(x_p)= \frac{1}{\sqrt{2^{a}(a!)\sqrt{\pi}}}\exp(-\frac{x^2_p}{2})H_a(x_p),
\end{equation}
where $H_a(x_p)$ are the physicist's Hermite polynomials. The phononic part of the overall wavefunction of a Fock state of a given electronic state is the product of the four quantum harmonic oscillator wavefunctions, each describing the uncoupled coordinates of vibration dictated by the normal modes of the same electronic state. In their respective natural units, they are,
\begin{equation}\label{eqS: phononic wavefunction}
    \psi^{\ket{o}}_N= \big[\psi^{\ket{o}}_{n_1}(v^{\ket{o}}_1\cdot r'')\big]\big[\psi^{\ket{o}}_{n_2}(v^{\ket{o}}_2\cdot r'')\big]\big[\psi^{\ket{o}}_{n_3}(v^{\ket{o}}_3 \cdot r'')\big]\big[\psi^{\ket{o}}_{n_4}(v^{\ket{o}}_4 \cdot r'')\big].
\end{equation}
Using Eq.~(\ref{eqS: phononic wavefunction}), the wavefunction overlap integral becomes
\begin{equation}
    S^{o_a,o_b}_{N,N'}= \int_{-\infty}^{\infty}\int_{-\infty}^{\infty} \int_{-\infty}^{\infty}\int_{-\infty}^{\infty} \psi^{\ket{o_a}}_N\exp(ik\zeta_l''\sqrt{\frac{\hbar}{m_{\rm a}\omega_{\rm t}}})\psi^{\ket{o_b}}_{N'}\,dz_{1}''\,dz_{2}''\,d\zeta_{1}''\,d\zeta_{2}''.
\end{equation}
\section{RAIM in a Paul trap Hamiltonian}
 \label{sec:Paul}
In this section, we include the contribution of the Paul trap to the total Hamiltonian for the two different orientations of the RAIMs (along the radial and rf-null axis) that were considered in the main text.
\subsection{RAIM oriented along the rf-null axis}
As shown in Fig.~\ref{fig:fig1}(a)(i) in the main text, the Paul trap's potential is in the $xy$-plane. Here, the Rydberg atom core and the ion are on the rf-null axis with the core's position defining the origin and the ion at $z=r_{\rm ci}$. The Rydberg electron is at $r_{\rm e}$. The Hamiltonian $H_{\rm P, ax}(r_{\rm ci}, t)$ is given by
\begin{equation}
  \label{eqS:HPaulax}
  \begin{split}
     H_{\rm P, ax}(r_{\rm ci}, t)&= \frac{m_{\rm i}q\Omega_{\rm rf}^2}{4}\big(y_{\rm e}^2-x_{\rm e}^2\big) \cos(\Omega_{\rm rf}t)\\
     &=\frac{m_{\rm i}q\Omega_{\rm rf}^2}{4}\big(r_{\rm e}^2\sin^2{\theta_{\rm e}}\sin^2{\phi_{\rm e}}-r_{\rm e}^2\sin^2{\theta_{\rm e}}\cos^2{\phi_{\rm e}}\big) \cos(\Omega_{\rm rf}t)\\
     &=\frac{m_{\rm i}q\Omega_{\rm rf}^2}{4}\big(-r_{\rm e}^2\sin^2{\theta_{\rm e}}\cos{2\phi_{\rm e}}\big) \cos(\Omega_{\rm rf}t),\\
  \end{split}
\end{equation}
where $m_{\rm i}$ and $\Omega_{\rm rf}$ are the mass of the trapped ion and the trap drive frequency respectively. $q$ ($=|q_x|=|q_y|$) is the dynamic stability parameter (measure of the electric field gradient in the trap) given by \cite{RevModPhys.75.281},
\begin{equation}
  q_{x,y}= \frac{2e V^{*}\alpha^{*}_{x,y}}{m_{\rm i} \Omega_{\rm rf}^2}
\end{equation}
where $\alpha^{*}_{x,y}$ are the trap's geometric factors and $V^{*}$ is the voltage at the RF electrodes. 
\subsection{RAIM oriented along the radial axis}
As shown in Fig.~\ref{fig:fig1}(a)(ii) in the main text, we consider a Paul trap potential in the $zy$-plane. Here, the Rydberg atom core and the ion are in the same $zy$-plane with the core's position defining the origin and the ion being on the rf null axis at, $z=r_{\rm ci}$. The Rydberg electron is at $r_{\rm e}$. The Hamiltonian $H_{\rm P, rad}(r_{\rm ci}, t)$ is given by
\begin{equation}
\label{eqS:HPaulrad}
  \begin{split}
     H_{\rm P, rad}(r_{\rm ci}, t)&= \frac{m_{\rm i}q\Omega_{\rm rf}^2}{4}\big(y_{\rm e}^2-(r_{\rm ci}-z_{\rm e})^2+r_{\rm ci}^2\big) \cos(\Omega_{\rm rf}t)\\
     &=\frac{m_{\rm i}q\Omega_{\rm rf}^2}{4}\big(y_{\rm e}^2-z_{\rm e}^2+2r_{\rm ci}z_{\rm e}\big) \cos(\Omega_{\rm rf}t)\\
     &=\frac{m_{\rm i}q\Omega_{\rm rf}^2}{4}\big(r_{\rm e}^2\sin^2{\theta_{\rm e}}\sin^2{\phi_{\rm e}}-r_{\rm e}^2\cos^2{\theta_{\rm e}}+2r_{\rm ci}r_{\rm e}\cos{\theta_{\rm e}}\big) \cos(\Omega_{\rm rf}t)\\
     &=\frac{m_{\rm i}q\Omega_{\rm rf}^2}{4}\biggl(\frac{r_{\rm e}^2\sin^2{\theta_{\rm e}}}{2}-\frac{r_{\rm e}^2\sin^2{\theta_{\rm e}}\cos{2\phi_{\rm e}}}{2}-r_{\rm e}^2\cos^2{\theta_{\rm e}}+2r_{\rm ci}r_{\rm e}\cos{\theta_{\rm e}}\biggr) \cos(\Omega_{\rm rf}t).
  \end{split}
\end{equation}
\section{Effect of the second ion and axial trap on RAIM}\label{sec:Effect of the second ion and axial trap on RAIM}
\begin{figure}
    \centering
\includegraphics{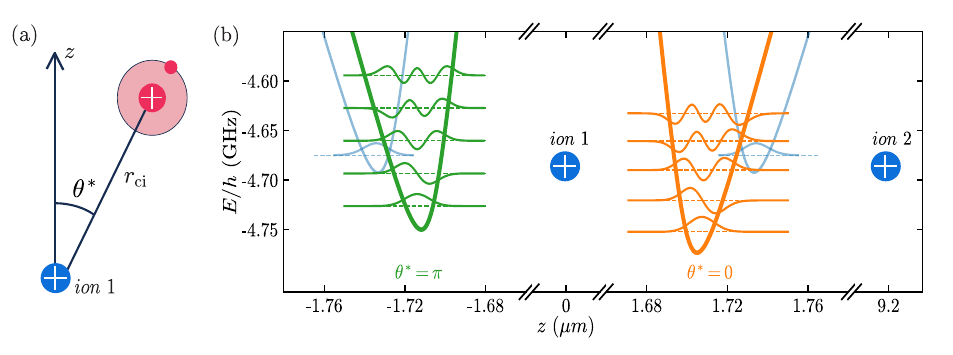}
    \caption{(a) Rydberg atom-ion molecule, where $\theta^*$ is the angle between the molecular axis and the $z$-axis. (b)  50$P_{1/2}$ molecular well state of a ${}^{87}{\rm Rb}{}^{9}{\rm Be}^+$ RAIM with its first five bound states after including the electric fields from the axial trap electrodes that traps ${}^{9}{\rm Be}^+$ ions at a trap frequency of $\omega_{\rm i}= 2\pi\times 1$ MHz and the second ${}^{9}{\rm Be}^+$ ion for two different orientations (orange curve for $\theta^*=0$ and green curve for $\theta^*=\pi$). Blue curves represent the well state under the influence of only the nearest ion. The energies are relative to the field-free atomic 50$P_{3/2}$ state.}
\label{fig:Rb87_50Pwell_2ionsztrap}
\end{figure}

In this section, we explain in detail the effects of the second ion and the axial trap electrodes on the RAIM for two possible orientations ($\theta^*=0, \pi$) along the rf-null axis ($z$-axis). We assume a crystal of two ${}^{9}{\rm Be}^+$ ions with an axial trap frequency $\omega_{\rm i}= 2\pi\times 1$ MHz, with ion 1 at the origin and ion 2 at $z=d_{12}$ ($\approx 9.2\mu$m) as shown in Fig.~\ref{fig:Rb87_50Pwell_2ionsztrap}(b). The axial trap's minima where the potential is zero is at $z=d_{12}/2$.

\subsection{When $\theta^*=\pi$}
In this case, a ${}^{87}{\rm Rb}$ Rydberg atom is placed on the left of the origin with its ionic core at $z=-r_{\rm ci}$. The overall Hamiltonian ${H}^{\pi}\left(r_{\rm ci}\right)$ is given by
\begin{equation}\label{eqS:Hl}
\begin{split}
     &{H}^{\pi}\left(r_{\rm ci}\right) = H_0 + H^{\pi}_{\rm ai,1}\left(r_{\rm ci}\right)+  H^{\pi}_{\rm ai,2}\left(r_{\rm ci}\right) +H^{\pi}_{\rm P, z}\left(r_{\rm ci}\right),\\
     &H^{\pi}_{\rm ai,1}\left(r_{\rm ci}\right)= -\frac{e^2}{4\pi\epsilon_0}\sum_{l^\prime=1}^{\infty}\sqrt{\frac{4\pi}{2l^\prime+1}}\frac{r_{\rm e}^{l^\prime}}{r_{\rm ci}^{l^\prime+1}}Y_{l^\prime0}(\theta_{{\rm e}},\phi_{{\rm e}}),\\
     &H^{\pi}_{\rm ai,2}\left(r_{\rm ci}\right)= -\frac{e^2}{4\pi\epsilon_0}\sum_{l^\prime=1}^{\infty}\sqrt{\frac{4\pi}{2l^\prime+1}}\frac{r_{\rm e}^{l^\prime}}{(r_{\rm ci}+d_{12})^{l^\prime+1}}Y_{l^\prime0}(\theta_{{\rm e}},\phi_{{\rm e}}),\\
     &H^{\pi}_{\rm P,z}\left(r_{\rm ci}\right)=\frac{m_{\rm i}\omega_{\rm i}^2}{2}\Big((d_{12}+2r_{\rm ci})r_{\rm e} \cos\theta_{\rm e}-r_{\rm e}^2 \cos^2\theta_{\rm e}\Big).
\end{split}
\end{equation}
\subsection{When $\theta^*=0$}
In this case, a ${}^{87}{\rm Rb}$ Rydberg atom is placed on the right of the origin with its ionic core at $z=r_{\rm ci}$. The overall Hamiltonian ${H}^{0}\left(r_{\rm ci}\right)$ is given by
\begin{equation}\label{eqS:Hr}
\begin{split}
     &{H}^{0}\left(r_{\rm ci}\right) = H_0 + H^0_{\rm ai,1}\left(r_{\rm ci}\right)+  H^0_{\rm ai,2}\left(r_{\rm ci}\right) +H^0_{\rm P, z}\left(r_{\rm ci}\right),\\
     &H^0_{\rm ai,1}\left(r_{\rm ci}\right)= -\frac{e^2}{4\pi\epsilon_0}\sum_{l^\prime=1}^{\infty}\sqrt{\frac{4\pi}{2l^\prime+1}}\frac{r_{\rm e}^{l^\prime}}{r_{\rm ci}^{l^\prime+1}}(-1)^{l^\prime}Y_{l^\prime0}(\theta_{{\rm e}},\phi_{{\rm e}}),\\
     &H^0_{\rm ai,2}\left(r_{\rm ci}\right)= -\frac{e^2}{4\pi\epsilon_0}\sum_{l^\prime=1}^{\infty}\sqrt{\frac{4\pi}{2l^\prime+1}}\frac{r_{\rm e}^{l^\prime}}{(-r_{\rm ci}+d_{12})^{l^\prime+1}}Y_{l^\prime0}(\theta_{{\rm e}},\phi_{{\rm e}}),\\
     &H^0_{\rm P,z}\left(r_{\rm ci}\right)=\frac{m_{\rm i}\omega_{\rm i}^2}{2}\Big((d_{12}-2r_{\rm ci})r_{\rm e} \cos\theta_{\rm e}-r_{\rm e}^2 \cos^2\theta_{\rm e}\Big).
\end{split}
\end{equation}

In both cases, we see that the electric field from the second ion and the axial trap electrodes displaces the well minima along the $z$-axis and the energy axis. At the same time, the well's shape too changes resulting in a slightly different RAIM natural frequency of $\omega^{\pi}_{\rm M}= 2\pi\times 33$ MHz compared to the value used ($\omega_{\rm M}= 2\pi\times 36$ MHz) for (anti-) blockade simulation. But, note that the overall physics that is being proposed shall remain the same, as the collective eigenmodes and eigenfrequencies upon using the new value of $\omega_{\rm M}$ changes negligibly. It should also be noted that the effect of the second ion's vibrations about its stable position on the RAIM's potential can be ignored here, due to the large second ion-atom separation and axial trap frequency.
\section{Scaling laws}
\label{sec:scaling}
\begin{figure*}[h]
    \centering
    \includegraphics{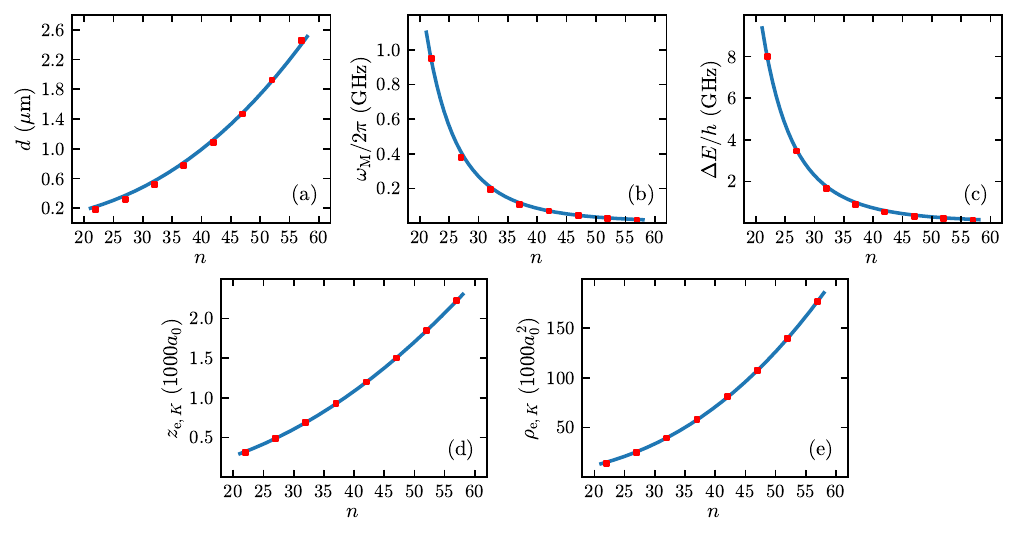}
    \caption{Properties related to the $nP_{1/2}$ molecular well of a ${}^{87}{\rm Rb}{}^{9}{\rm Be}^+$ RAIM as a function of the principal quantum number $n$ with its corresponding fits (blue). (a) Position $d$. (b) Harmonic frequency component $\omega_M$. (c) Energy gap $\Delta E$. (d) Transition dipole moment $z_{{\rm e},K}$. (e) Transition quadrupole moment $\rho_{{\rm e},K}$.}
    \label{fig:Scaling laws}
\end{figure*}
In this section, we discuss the scaling laws that were used in the main text, specifically the atom-ion distance $d$, the molecular potential harmonic frequency $\omega_{\rm M}$, the energy gap between the well state and the nearest (in energy) Rydberg state $K$ at the location of the well minima ($r_{\rm ci}=d$) $\Delta E$, the transition dipole moment $z_{{\rm e},K}$ for RAIMs oriented along the radial axis, and the quadrupole moment $\rho_{{\rm e},K}=y_{{\rm e},K}^2-x_{{\rm e},K}^2$ for RAIMs oriented along the rf-null axis.

By numerically fitting the values of $d$ and the harmonic frequency component ($\omega_{\rm M}$) for the $P_{1/2}$ molecular wells of a ${}^{87}{\rm Rb}{}^{9}{\rm Be}^+$ RAIM at different $n$, we extract the corresponding scaling laws, see Fig.~\ref{fig:Scaling laws}(a), (b) and (c). Specifically, we find $d\approx 1.85 a_0 n^{5/2}$ (whose scaling can also be derived from the Inglis Teller limit as in the main text), $\omega_{\rm M}\approx \omega_0 n^{-4}$, with $\omega_0= 2\pi\times 218101$ GHz and $\Delta E\approx0.28 E_{\rm H}/n^{4}$, with $E_{\rm H}$ the Hartree energy, respectively. 

When the RAIM is formed along one of the trap's radial axes, its decay is dominated by the coupling via the transition dipole moment ($z_{{\rm e},K}$) in the last term in Eq.~(\ref{eqS:HPaulrad}), which only couples states with the same value of $m_j$ (which in our case is +1/2). As mentioned in the main text, we focus only on the $P_{1/2}$ well state's coupling to the nearest (in energy) atomic Rydberg state $K$. We find two nearly degenerate states with couplings $z_1$ and $z_2$, so we consider the transition dipole moment ($z_{{\rm e},K}=\sqrt{z_1^2+z_2^2}$) to account for this degeneracy. By numerically fitting the value of $z_{{\rm e},K}$ obtained for different values of $n$, we get an expression for $z_{{\rm e},K}\approx 0.6a_0n^2$, see Fig.~\ref{fig:Scaling laws}(d). When the RAIM is formed along the rf-null axis, its decay is dictated by the coupling via the transition quadrupole moment $\rho_{{\rm e},K}$ to the nearest (in energy) atomic Rydberg states in the $m_j=-3/2,+5/2$ spectrum (i.e.,~$\Delta m_j=2$). Here, we focus on the pair of two nearly degenerate states (one from the $m_j=-3/2$ spectrum and the other from the $m_j=+5/2$ spectrum) with couplings $\rho_1$ and $\rho_2$, and calculate the root of their square's sum ($\rho_{{\rm e},K}=\sqrt{\rho_1^2+\rho_2^2}$). We ignore the nearest level (one of the states in the $m_j=-3/2$ spectrum) due to its small $\rho_{{\rm e},K}$ value, although we remark that for $n \gtrsim 150$ this transition starts to dominate since it is described by a different power law. By numerically fitting the value of $\rho_{{\rm e},K}$ obtained for different values of $n$, we find $\rho_{{\rm e},K}\approx 4.8a_0^2n^{2.6}$, see Fig.~\ref{fig:Scaling laws}(e).

\section{RAIMs in a digital Paul trap}
\label{sec:digital}
\begin{figure*}[h]
    \centering 
\includegraphics[width=7in,height=1.8in]{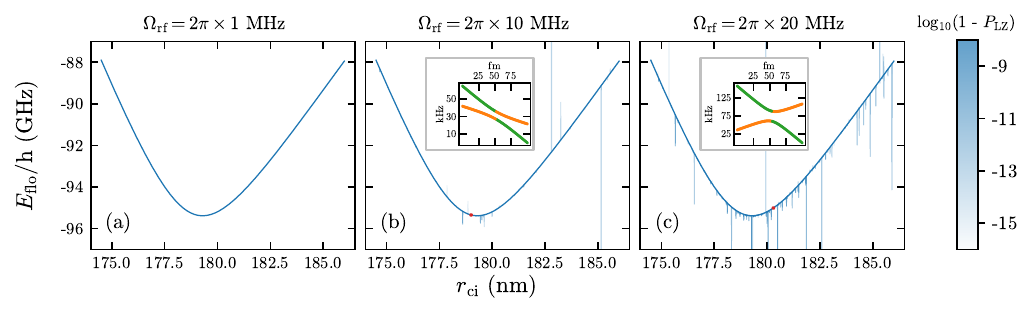}
    \caption{Energy expectation value of the static Hamiltonian $H_{\rm TI}$ for the Floquet state $\ket{\phi_{\rm RAIM}}$ ($E_{\rm flo}$) that correspond to the 22$P_{1/2}$ molecular well state of a ${}^{87}{\rm Rb}{}^{9}{\rm Be}^+$ RAIM oriented along one of the radial axis of a 2 state-digital Paul trap which traps ${}^{9}{\rm Be}^+$ ions with $q = 0.1$. Energies are relative to the field-free atomic 22${P}_{3/2}$ state. The scan over $r_{\rm ci}$ was done at a resolution of 10 pm. The spikes are shaded according to $1-P_{\rm LZ}$ at each avoided crossing. (a), (b) and (c) correspond to different drive frequencies 2$\pi$ $\times$ 1 MHz, 2$\pi$ $\times$ 10 MHz, and 2$\pi$ $\times$ 20 MHz respectively. The insets show the avoided crossings between the quasienergy of the Floquet state that correspond to the well state (orange) and to the other eigenstate (grey) it couples with at the core-ion distance ($r_{\rm ci}$) marked by the red dot in the main panels.}
    \label{fig:Rb87 22Phalf_2SDIT_avoided crossings transparent}
\end{figure*}

In this section, we discuss the effect of a 2-state digital Paul trap on RAIMs. The same procedure discussed in the main text is used to evaluate $E_{\rm flo}$ that correspond to the well state of an RAIM oriented along the radial axis of a trap driven at different trap frequencies as shown in Fig. ~\ref{fig:Rb87 22Phalf_2SDIT_avoided crossings transparent}. The increased number of spikes (avoided crossings) here compared to the sinusoidal case is due to the spectrally broadened nature of the input drive that results in resonant (albeit weak) direct single-photon couplings to many Rydberg states, which simultaneously results in a larger number of pathways through which high-order photon processes couple different Rydberg states. 

We note that the probability of a diabatic transition at each avoided crossing in the quasienergy spectrum (denoted by the spikes in Fig.~\ref{fig:Rb87 22Phalf_2SDIT_avoided crossings transparent}) is given by the Landau-Zener criterion $P_{\rm LZ}=\exp(-2\pi\Gamma)$ where  
\begin{equation}\label{eq: PLZ}
   \Gamma=\frac{g^2/\hbar}{\left\vert \frac{\partial\Delta \epsilon_{\alpha}}{\partial r_{\rm ci}}\frac{\partial r_{\rm ci}}{\partial t}\right\vert}.
\end{equation}
Here, $g$ represents the coupling between the two states which can be estimated as half of the quasienergy gap at the avoided crossing. The first term in the denominator $\frac{\partial\Delta \epsilon_{\alpha}}{\partial r_{\rm ci}}$ is the rate at which the quasienergy gap between the two states change as a function of $r_{\rm ci}$, far away from resonance. We approximate the other term in the denominator as $\frac{\partial r_{\rm ci}}{\partial t}\approx\omega_{\rm M} l_{\rm ho}$, where $l_{\rm ho}=\sqrt{\hbar/\left(2\mu_{} \omega_{\rm M}\right)}$ is the characteristic length scale of the oscillations and $\mu_{}$ is the reduced mass of the RAIM. 

\section{Details on Anti-Blockade Pulse Optimization}
\label{sec:pulse}
To realize anti-blockade (facilitation) and avoid the population collecting in the intermediate states $\ket{gR}$ and $\ket{Rg}$, we modulate the two-tone Rabi frequency $\Omega_L = \Omega_1 f_1(t) e^{i \omega_1 t} + \Omega_2 f_2(t) e^{i \omega_2 t}$, where $f_{1,2}(t)$ are the pulse shapes to be optimized. We use a pulse shape inspired by the chopped random-basis (CRAB) algorithm~\cite{PhysRevLett.106.190501} to optimize the frequency components of a correction to a Gaussian pulse. The pulse shape is of the form
\begin{align}
    f_i(t) = \left| A_i e^{-(t-b_i)^2/a_i}\left[1 + \sum_{j=1}^{N_p} c_{i,j} \cos( \omega_{i,j} t) \right] \right|, \label{eq:pulseoptimisation}
\end{align}
where $a_i,b_i$ determine the length and location of the Gaussian pulse, with $c_{i,j},\omega_{i,j}$ optimisation parameters and $N_p$ the number of frequency components. Here $A_i$ is a normalisation factor to ensure that ${\rm max}_{0\leq t \leq t_{\rm f}}[f_{1,2}(t)] = \Omega_{1,2}$ for each drive $\Omega_1$ and $\Omega_2$.

We perform a two-step optimization. Firstly, we set $a_i$ and $b_i$ to minimize the population in $\ket{gR}$ and $\ket{Rg}$ at the end of the pulse sequence. Secondly, we perform a random search to minimize the cost function
\begin{align}
    C = |\abs{\braket{\psi(t_f)}{gg}}^2-0.5| + \abs{\braket{\psi(t_f)}{gR}}^2 +|\abs{\braket{\psi(t_f)}{RR}}^2-0.5|,
\end{align}
which penalizes the population of $\ket{gR}$ whilst encouraging the equal population of $\ket{gg}$ and $\ket{RR}$. We find that optimizing this simple cost function with $N_p = 3$ frequency components converges relatively quickly to an acceptable solution (i.e., low infidelity and within short times $t_f \lesssim 10 \mu \text{s}$) using only $\sim 200$ random parameter draws. The results of the optimization are shown in Fig.~\ref{fig:blockadefacilitation} of the main text. Finally, we note that the use of more sophisticated optimization algorithms, for example local gradient-based or probabilistic algorithms, may unveil pulse shapes that yield higher fidelity facilitation in shorter pulse times. 

\putbib[suppbib]
\end{bibunit}

\end{document}